\documentclass[twocolumn,english,prx,showpacs]{revtex4-1}
\usepackage[colorlinks=true,urlcolor=blue,citecolor=blue,linkcolor=blue]{hyperref}
\usepackage[T1]{fontenc}
\usepackage[utf8]{inputenc}
\usepackage{amssymb}
\usepackage{graphicx}
\usepackage{amsmath,color}
\usepackage{mathrsfs}
\usepackage{float}
\usepackage{indentfirst}
\usepackage{braket}
\usepackage{babel}
\usepackage{bbold}
\usepackage{booktabs}
\usepackage{multirow}
\usepackage{pbox}

\makeatletter

\def\eqa{\begin{eqnarray}}
\def\eea{\end{eqnarray}}
\newcommand{\eq}{\begin{equation}}
\newcommand{\ee}{\end{equation}}

\newcommand{\Eq}[1]{Eq.~(\ref{#1})}

\newcommand{\Tr}{{\rm Tr}}

\makeatother


\begin{document}

\title{Fidelity susceptibility made simple: \\A unified quantum Monte Carlo approach}

\author{Lei Wang$^{1}$, Ye-Hua Liu$^{1}$, Jakub Imri\v{s}ka$^{1}$, Ping Nang Ma$^{2}$ and Matthias Troyer$^{1}$}

\affiliation{$^{1}$Theoretische Physik, ETH Zurich, 8093 Zurich, Switzerland}
\affiliation{$^{2}$Yotcopi Technologies, National University of Singapore}

\begin{abstract}
The fidelity susceptibility is a general purpose probe of phase transitions. With its origin in quantum information  and in the differential geometry perspective of quantum states, the fidelity susceptibility can indicate the presence of a phase transition without prior knowledge of the local order parameter, as well as reveal the universal properties of a critical point. The wide applicability of the fidelity susceptibility to quantum many-body systems is, however, hindered by the limited computational tools to evaluate it. We present a generic, efficient, and elegant approach to compute the fidelity susceptibility of correlated fermions, bosons, and quantum spin systems in a broad range of quantum Monte Carlo methods. It can be applied both to the ground-state and non-zero temperature cases. The Monte Carlo estimator has a simple yet universal form, which can be efficiently evaluated in simulations. We demonstrate the power of this approach with applications to the Bose-Hubbard model, the spin-$1/2$ XXZ model, and use it to examine the hypothetical intermediate spin-liquid phase in the Hubbard model on the honeycomb lattice.
\end{abstract}
\maketitle

\tableofcontents

\section{Introduction}

Phase transitions highlight the beauty of universality, despite the great diversity of nature. For example, one finds a unified description for systems ranging from ultracold bosons~\cite{Fisher:1989vw, Greiner:2002wt} to magnetic insulators~\cite{PhysRevLett.100.205701, Giamarchi:2008wu, Coldea:2010is} on the verge of a phase transition. Phase transitions origin from the competition between different tendencies when a macroscopic system tries to organize itself. Thermal fluctuations can drive classical phase transitions at non-zero temperatures, while quantum phase transitions can occur even at zero-temperature because of the competition between non-commuting terms in the quantum mechanical Hamiltonian~\cite{Hertz:1976ta, Sachdevbook}. At the phase transition point, physical observables often exhibit singular behavior. In this respect, phase transitions are the most dramatic manifestation of the laws of statistical and quantum mechanics.

Traditional descriptions of phase transitions are based on low-energy effective theories of local order parameters, which have had enormous success in explaining various phase transitions of superfluids, superconductors~\cite{gorkov1959microscopic}, and quantum magnets~\cite{PhysRevLett.60.1057,PhysRevLett.71.169}. However, in recent years, exceptions to this Ginzburg--Landau--Wilson paradigm have emerged~\cite{WenBook}. In particular, topological phase transitions~\cite{PhysRevLett.70.1501, Read:2000vz, Kitaev:2006ik} do not have a local order parameters on either side of the phase transition. Therefore, new theoretical tools are needed to search for and characterize these new quantum phases and phase transitions. Many concepts in quantum information science~\cite{QCQIBook}, such as the quantum fidelity and quantum entanglement, have proven to be useful~\cite{Gu:2010em, Eisert:2010hq}. Having a point of view which is totally different from the traditional condensed matter approach, they do not assume the presence of a local order parameter and thus offer new perspectives of the phase transitions and their universalities.

Specifically, we consider the following one-parameter family of Hamiltonians with a driving parameter $\lambda$, 

\begin{equation}
\hat{H}(\lambda) = \hat{H}_{0} + \lambda \hat{H}_{1}. 
\label{eq:split}
\end{equation}
As $\lambda$ changes, the system may go through one or several phase transition(s) because of the competition between $\hat{H}_{0}$ and $\hat{H}_{1}$.
The \emph{quantum fidelity} measures the distance on the manifold of $\lambda$, which is defined as the overlap between the ground-state wavefunctions at two different values of the driving parameter,

\begin{equation}
F(\lambda,\epsilon) = {\left|\braket{\Psi_{0}(\lambda)|\Psi_{0}(\lambda+\epsilon)}\right|},
\label{eq:gsfidelity}
\end{equation}
where $\hat{H}(\lambda) \ket{\Psi_{n}(\lambda)}= E_{n}(\lambda)\ket{\Psi_{n}(\lambda)}$ and $n=0$ corresponds to the ground state. Unless otherwise stated, we assume the wavefunctions are normalized and there is no ground-state degeneracy. It is anticipated that the fidelity will exhibit a dip when the two wavefunctions are qualitatively different, e.g., when they belong to different phases~\cite{Zanardi:2006fba}. This wavefunction overlap is also related to the Anderson's orthogonality catastrophe~\cite{PhysRevLett.18.1049} and the Loschmidt echo in quantum dynamics~\cite{Quan:2006fda}.

Since in general the quantum fidelity vanishes exponentially with the system size for a many-body system, it is more convenient to study the change of its logarithm with respect to the driving parameter, called the \emph{fidelity susceptibility}~\cite{You:2007ew},
\begin{equation}
\chi_{F}(\lambda) = -\frac{\partial^{2} \ln F}{\partial{\epsilon}^{2}} \bigg|_{\epsilon=0}.
\label{eq:definition}
\end{equation}
%
The first-order derivative vanishes because $F$ is at its maximum when $\epsilon=0$. In general, the fidelity susceptibility is an extensive quantity away from the critical point, but it exhibits a maximum or even diverges at the critical point, thus indicating a quantum phase transition~\cite{CamposVenuti:2007il, Gu:2009kg}. Similar to conventional thermodynamic quantities, it also follows a scaling law close to the critical point~\cite{CamposVenuti:2007il, Gu:2009kg, Schwandt:2009jl, Albuquerque:2010fv}, which can be used to extract universal information about the phase transition. An important feature of the fidelity susceptibility is that it can reveal a phase transition \emph{without} prior knowledge of the local order parameter. This makes it suitable for detection of topological phase transitions~\cite{Abasto:2008gh,Yang:2008cl,Zhao:2009hc, PhysRevA.79.032302} and for tackling challenging cases where an in-depth understanding of the underlying physics is still lacking~\cite{PhysRevB.80.094529, PhysRevB.84.125113}. Interestingly, the fidelity susceptibility may also be accessible to experiments~\cite{Zhang:2008iga,Kolodrubetz:2013bfa, Gu:2014gk}.

Despite its appealing features, the difficulty in calculating the fidelity susceptibility has hindered its use in numerical simulations. 
Many previous studies were thus limited to the cases where the ground-state wavefunction overlap could be calculated from the analytical solution, exact diagonalization, or density-matrix-renormalization-group (DMRG) methods~\cite{Gu:2010em}.

There are several equivalent formulations of the fidelity susceptibility~\Eq{eq:definition}, which reveal different  aspects of the quantity. From a computational point of view, they offer direct ways to calculate the fidelity susceptibility without the need to perform numerical derivatives of the fidelity as in \Eq{eq:gsfidelity}.

(a)~Expanding $\ket{\Psi_{0}(\lambda + \epsilon)}$  for small $\epsilon$, one can cast the definition Eq.~(\ref{eq:definition}) into an explicit form~\cite{Zanardi:2007bea, CamposVenuti:2007il},

\begin{equation}
\chi_{F}(\lambda)  =  \frac{ \braket{\partial_{\lambda} \Psi_{0} |  \partial_{\lambda} \Psi_{0}} }{ \braket{ \Psi_{0}|  \Psi_{0} }} - \frac{  \braket{ \Psi_{0} | \partial_{\lambda} \Psi_{0} }
  }{\braket{\Psi_{0}  |  \Psi_{0} }}  \frac{ \braket{ \partial_{\lambda}
  \Psi_{0}| \Psi_{0}} }{  \braket{\Psi_{0}| \Psi_{0}}}.
  \label{eq:differentialform}
\end{equation}
The above form does not assume properly normalized wavefunctions $\ket{\Psi_{0}}$. Equation~(\ref{eq:differentialform}) reveals the geometric content of the fidelity susceptibility~\cite{CamposVenuti:2007il,Zanardi:2007bea}, since this expression is the real part of the quantum geometric tensor~\cite{Provost:1980ut}.

(b)~Alternatively, one can calculate the first-order perturbation for $\ket{\Psi_{0}(\lambda + \epsilon)}$ and get
\begin{equation}
\chi_{F}(\lambda) = \sum_{n\neq 0} \frac{|\braket{ \Psi_{n}(\lambda) |\hat{H}_{1}| \Psi_{0}(\lambda)}  |^{2}}{\left[E_{n}(\lambda)-E_{0}(\lambda)\right]^{2}}.
\label{eq:perturbationform}
\end{equation}
Compared to \Eq{eq:differentialform}, \Eq{eq:perturbationform} does not contain derivatives but involves all eigenstates and the full spectrum. It explicitly shows that $\chi_{F}\ge0$ and suggests the divergence of $\chi_{F}$ when the energy gap of the system closes.

(c)~Reference~\cite{You:2007ew} views \Eq{eq:perturbationform} as the zero-frequency component of a spectral representation, thus a Fourier transform is performed to obtain an alternative expression,

\begin{align}
  \chi_F (\lambda)= \int_0^{\infty} d\tau\, \left[\braket{\Psi_{0}|\hat{H}_{1} \left({\tau} \right) \hat{H}_{1} |\Psi_{0} } -  \braket{\Psi_{0}| \hat{H}_{1}|\Psi_{0} }^2\right] \tau ,
  \label{eq:kuboform}
\end{align}
where $\hat{H}_{1}(\tau) =e^{\hat{H}\tau} \hat{H}_{1}e^{-\hat{H}\tau}$. Equation~(\ref{eq:kuboform}) has the form of a linear-response formula and is computationally more friendly than \Eq{eq:differentialform} or \Eq{eq:perturbationform}.
References~\cite{Schwandt:2009jl, Albuquerque:2010fv} generalize it to non-zero temperature by replacing the integration limit with $\beta/2$,

\begin{eqnarray}
  \chi_F(\lambda) & = & \int_0^{\beta/2} d\tau\, \left[\braket{\hat{H}_{1} \left({\tau} \right) \hat{H}_{1}} -  \braket{\hat{H}_{1}}^2\right] \tau,
  \label{eq:kubo}
\end{eqnarray}
where $\braket{\ldots}$ denotes the thermal average at inverse temperature $\beta$. Besides reducing to \Eq{eq:kuboform} as $\beta\rightarrow\infty$, \Eq{eq:kubo} is nevertheless a well-defined quantity at nonzero temperatures. It bounds the divergence of an alternative ``mixed state'' fidelity susceptibility~\cite{Zanardi:2007ir,Zanardi:2007il}, which is based on the Uhlmann fidelity~\cite{Uhlmann:1976ti, Jozsa:1994ik}, and both quantities follow the same scaling law close to a quantum critical point~\cite{Schwandt:2009jl, Albuquerque:2010fv}. In general, the evaluation of Eq.~(\ref{eq:kubo}) is still a formidable computational task, which requires \emph{ad hoc} implementation depending on the details of the Hamiltonian. For example, the fidelity susceptibility for two-dimensional quantum spin systems is calculated in Refs.~\cite{Schwandt:2009jl, Albuquerque:2010fv} using quantum Monte Carlo method; while for a one-dimensional quantum spin system, Ref.~\cite{Sirker:2010fu} computed it using transfer-matrix DMRG method.
The nontrivial implementation of these specific approaches and the overhead in the calculation still limits the wide applicability of the fidelity susceptibility approach to a broad range of quantum many-body systems.

In this paper we present a simple yet generic approach to compute the fidelity susceptibility in a large variety of modern quantum Monte Carlo methods, including the continuous-time worldline~\cite{PhysRevLett.77.5130,Prokofev:1998tc,Evertz:2003ch,Kawashima:2004clb} and stochastic series expansion (SSE)~\cite{Sandvik:1991tn} methods for bosons and quantum spins, and the diagrammatic determinantal methods for quantum impurity~\cite{Rubtsov:2005iw, Werner:2006ko, Gull:2008cm, Gull:2011jd} and fermion lattice models~\cite{1999PhRvL..82.4155R, Burovski:2006hv, Iazzi:2014vv}. 
In all cases, the Monte Carlo estimator is generic and the implementations are straightforward. As long as the quantum Monte Carlo simulation is feasible (not hindered by the sign problem), the fidelity susceptibility can be easily calculated. Our finding can boost the investigation of quantum phase transitions from a quantum information perspective and becomes especially advantageous for the exploration of exotic phases beyond the Ginzburg--Landau--Wilson paradigm.


The organization of the paper is as follows. In Sec.~\ref{sec:estimator} we present our estimator for the fidelity susceptibility, and discuss its implementations in various quantum Monte Carlo methods. Section~\ref{sec:proof} presents derivations of the estimator. In Sec.~\ref{sec:application} we demonstrate the power of the fidelity susceptibility approach with applications to various models, including correlated bosons, fermions, and quantum spins, using a variety of quantum Monte Carlo methods.
Section~\ref{sec:discussion} discusses the relation between the zero-temperature and non-zero temperature estimators for the fidelity susceptibility and compares them to the previous approaches~\cite{Schwandt:2009jl, Albuquerque:2010fv}. We conclude with future prospects in Sec.~\ref{sec:outlook}.

\section{Results~\label{sec:estimator}}

We first present our results on the estimator of the fidelity susceptibility in a general setting, then discuss its implementations in various QMC methods, including the continuous-time worldline~\cite{PhysRevLett.77.5130,Prokofev:1998tc,Evertz:2003ch,Kawashima:2004clb} and diagrammatic determinantal approaches~\cite{Rubtsov:2005iw, Werner:2006ko, Gull:2008cm, Gull:2011jd, 1999PhRvL..82.4155R, Burovski:2006hv, Iazzi:2014vv}, and the stochastic series expansion method~\cite{Sandvik:1991tn}. In all cases, the fidelity susceptibility can be measured with little effort.

\subsection{Universal Covariance Estimator}

\begin{figure}[t]
 \centering
\includegraphics[width=8cm]{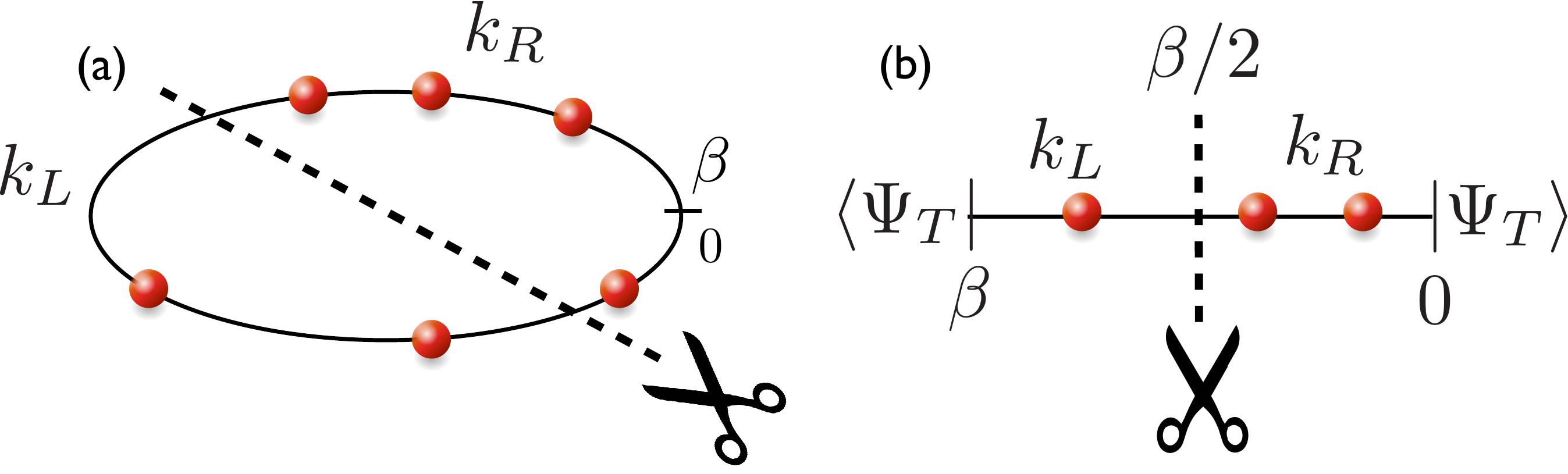}
\caption{Measurement of the fidelity susceptibility in the (a) non-zero temperature formalism and in the (b) ground-state projection scheme. Each red object represents a term in the driving Hamiltonian $\lambda\hat{H}_{1}$, denoted as a vertex. To measure the fidelity susceptibility Eqs.~(\ref{eq:finiteTkLkR},\ref{eq:zeroTkLkR}), we divide the imaginary-time axis into two halves and count the number of vertices $k_{L}$ and $k_{R}$ respectively. The non-zero temperature formalism allows an arbitrary division because of the \emph{periodic boundary condition} in the imaginary-time axis, while in the ground-state projection scheme the division has to be at $\beta/2$.}
\label{fig:concept}
\end{figure}

Many modern QMC methods~\cite{Sandvik:1991tn, PhysRevLett.77.5130, Prokofev:1998tc, 1999PhRvL..82.4155R, Evertz:2003ch, Kawashima:2004clb, Rubtsov:2005iw, Werner:2006ko, Burovski:2006hv, Gull:2008cm, Gull:2011jd, Iazzi:2014vv} share a unified conceptual framework, namely that the partition function is calculated as a perturbative series expansion,

\begin{equation}
Z = \Tr\left(e^{-\beta \hat{H}}\right) = \sum_{k=0}^{\infty} \lambda^{k} \sum_{\mathcal{C}_{k}}  w(\mathcal{C}_{k}),
\label{eq:ftexpansion}
\end{equation}
where the second summation runs over all the Monte Carlo configurations of a given expansion order $k$. The detailed meaning of the configuration depends on the specific QMC algorithm and will be explained in the next subsection. Figure~\ref{fig:concept}(a) depicts a generic configuration, where the $k$ objects residing on the periodic imaginary-time axis represent the vertices $\lambda\hat{H}_1$ in the expansion, with a Monte Carlo weight $\lambda^{k}w(\mathcal{C}_{k})$ for this configuration. QMC simulations~\cite{Sandvik:1991tn, PhysRevLett.77.5130, Prokofev:1998tc, 1999PhRvL..82.4155R, Evertz:2003ch, Kawashima:2004clb, Rubtsov:2005iw, Werner:2006ko, Burovski:2006hv, Gull:2008cm, Gull:2011jd, Iazzi:2014vv} sample the summation over $k$ and $\mathcal{C}_{k}$ on an equal footing. Specific algorithms differ by the detailed form of $w(\mathcal{C}_{k})$ and by the sampling schemes. Nevertheless, these QMC methods share a unified framework provided by Eq.~(\ref{eq:ftexpansion}), which is the only requirement for the estimator of the fidelity susceptibility \Eq{eq:kubo} to possess an appealing universal form in non-zero temperature QMC simulations,

\begin{eqnarray}
\chi^{T\ne 0}_{F}= \frac{\braket{k_{L} k_{R}} - \braket{k_{L}} \braket{k_{R}}}{2\lambda^{2}},  \label{eq:finiteTkLkR}
\end{eqnarray}
where $k_{L}$ and $k_{R}$ are the number of vertices residing in the range $[\beta/2, \beta)$ and $[0, \beta/2)$ of the imaginary-time axis, respectively, shown in Fig.~\ref{fig:concept}(a). In practice, however, because of the periodic boundary condition on the imaginary-time axis, the division of the time axis to halves may be done at an arbitrary location. Moreover, it is even possible to perform multiple measurements on the same configuration by generating several random divisions.

QMC methods~\cite{Sandvik:1991tn, PhysRevLett.77.5130, Prokofev:1998tc, 1999PhRvL..82.4155R, Evertz:2003ch, Kawashima:2004clb, Rubtsov:2005iw, Werner:2006ko, Burovski:2006hv, Gull:2008cm, Gull:2011jd, Iazzi:2014vv} can also be utilized at zero temperature, where the unnormalized ground-state wavefunction is obtained from an imaginary-time projection
\begin{equation}
\ket{\Psi_{0}} = \lim_{\beta\rightarrow\infty} e^{-\beta\hat{H}/2}\ket{\Psi_{T}}.
\label{eq:gs}
\end{equation}
Here $\beta$ is a projection parameter and the trial wavefunciton $\ket{\Psi_{T}}$ shall not be orthogonal to the true ground state. A similar framework as \Eq{eq:ftexpansion} applies, except that one now samples from the overlap $\bra{\Psi_{T}} e^{-\beta \hat{H}} \ket{\Psi_{T}}$ instead of the partition function.
In the projection scheme, the fidelity susceptibility has the estimator
\begin{equation}
\chi^{T=0}_{F} = \frac{\braket{k_{L} k_{R}}-  \braket{k_{L}} \braket{k_{R}}}{\lambda^{2}},
\label{eq:zeroTkLkR}
\end{equation}
where $k_{L}$ and $k_{R}$ are the number of vertices for the bra and ket states, which reside in the range $[\beta/2, \beta)$ and $[0, \beta/2)$ of the imaginary-time axis respectively, shown in Fig.~\ref{fig:concept}(b). Since the fidelity susceptibility is non-negative, the covariance formula \Eq{eq:zeroTkLkR} reveals positive correlation of $k_{L}$ and $k_{R}$ in a Monte Carlo simulation.


Equation~(\ref{eq:finiteTkLkR}) and \Eq{eq:zeroTkLkR} are the central results of the paper. As is obvious from the discussions in this section, neither details of the Hamiltonian, nor the statistics of the system need to be specified. These estimators are thus general and can be readily implemented in a variety of QMC methods for correlated fermionic, bosonic, or quantum spin systems~\cite{Sandvik:1991tn, PhysRevLett.77.5130, Prokofev:1998tc, 1999PhRvL..82.4155R, Evertz:2003ch, Kawashima:2004clb, Rubtsov:2005iw, Werner:2006ko, Burovski:2006hv, Gull:2008cm, Gull:2011jd, Iazzi:2014vv}. 

\subsection{Implementations}
We now discuss implementation of the estimators Eq.~(\ref{eq:finiteTkLkR}) and Eq.~(\ref{eq:zeroTkLkR}) in various concrete QMC methods. See Sec.~\ref{sec:ctqmc} for 
discussions about continuous-time worldline~\cite{PhysRevLett.77.5130,Prokofev:1998tc,Evertz:2003ch,Kawashima:2004clb} and diagrammatic determinantal~\cite{Rubtsov:2005iw, Werner:2006ko, Gull:2008cm, Gull:2011jd, 1999PhRvL..82.4155R, Burovski:2006hv, Iazzi:2014vv} approaches, and Sec.~\ref{sec:sse} for discussions about stochastic series expansion approach~\cite{Sandvik:1991tn}.

\subsubsection{Continuous-time worldline and diagrammatic determinantal approaches \label{sec:ctqmc}}

\begin{figure}[t]
 \centering
\includegraphics[width=8cm]{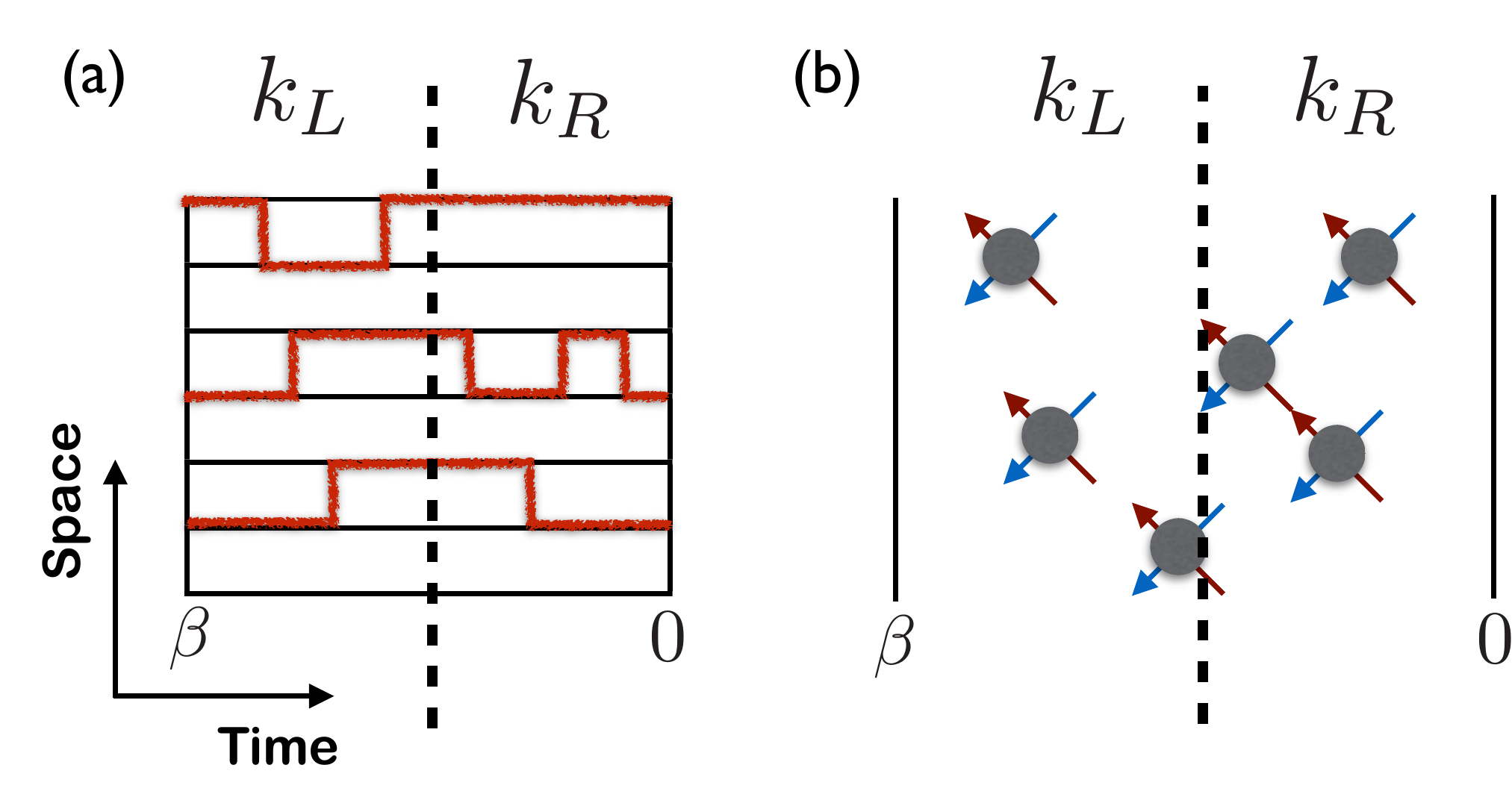}
\caption{(a) Measurement of the fidelity susceptibility in a continuous-time worldline QMC simulation of bosons and quantum spins, where one counts the number kinks $k_{L}$ and $k_{R}$ after division of the imaginary-time axis.
(b) In the diagrammatic determinantal QMC simulation of correlated fermions, the number of interaction vertices $k_{L}$ and $k_{R}$ are counted.}
\label{fig:CTQMC}
\end{figure}

Continuous-time worldline methods~\cite{PhysRevLett.77.5130,Prokofev:1998tc,Evertz:2003ch,Kawashima:2004clb} are widely used to simulate boson and quantum spin systems, while the diagrammatic determinantal approaches are the state-of-the-art methods for solving quantum impurity~\cite{Rubtsov:2005iw, Werner:2006ko, Gull:2008cm, Gull:2011jd} and fermion lattice models~\cite{1999PhRvL..82.4155R, Burovski:2006hv, Iazzi:2014vv}. A common feature of these methods is to split the Hamiltonian in the form of Eq.~(\ref{eq:split}) and perform a time-dependent expansion in $\lambda\hat{H}_{1}$,

\begin{eqnarray}
Z & =& \sum_{k=0}^{\infty} \lambda^{k} \int_{0}^{\beta} d\tau_{1}\ldots\int_{\tau_{k-1}}^{\beta} d\tau_{k}  \times \nonumber \\ & & \Tr\left[(-1)^{k} e^{-(\beta-\tau_{k})\hat{H}_{0}}\hat{H}_{1} \ldots  \hat{H}_{1}e^{ -\tau_{1}\hat{H}_{0} } \right], \label{eq:ct-expansion}
\end{eqnarray}
which obviously fits in the general framework of Eq.~(\ref{eq:ftexpansion}). In the continuous-time worldline approach~\cite{PhysRevLett.77.5130,Prokofev:1998tc,Evertz:2003ch,Kawashima:2004clb}, the $\hat{H}_{1}$ term corresponds to hoppings of bosons or spin flips, depicted as kinks of the worldlines in Fig.~\ref{fig:CTQMC}(a).  In continuous-time diagrammatic determinantal approaches~\cite{Rubtsov:2005iw, Werner:2006ko, Gull:2008cm, Gull:2011jd, 1999PhRvL..82.4155R, Burovski:2006hv, Iazzi:2014vv}, $\lambda\hat{H}_{1}$ contains the fermion interactions, drawn as interaction vertices in Fig.~\ref{fig:CTQMC}(b). 
Equation~(\ref{eq:ct-expansion}) has the form of a grand canonical partition function for a classical gas, where $\lambda$ plays the role of fugacity and $k$ is the number of certain classical objects (kinks or vertices) residing on the imaginary-time axis. Typical updates of continuous-time diagrammatic determinantal approaches~\cite{Rubtsov:2005iw, Werner:2006ko, Gull:2008cm, Gull:2011jd, 1999PhRvL..82.4155R, Burovski:2006hv, Iazzi:2014vv} consists of randomly inserting or removing vertices, which are identical to the updates of \emph{grand canonical Monte Carlo} method for molecular simulations~\cite{GCMC,MSBook}. For bosons and quantum spins there are more effective non-local updates such as the worm and directed loop updates~\cite{Prokofev:1998tc, Sandvik:1991tn, PhysRevLett.77.5130, Kawashima:2004clb, Evertz:2003ch}. In any case, the Monte Carlo estimators Eqs.~(\ref{eq:finiteTkLkR},\ref{eq:zeroTkLkR}) are independent to the detailed sampling procedures. It suffices to count $k_{L}$ and $k_{R}$ of Monte Carlo configurations to calculate the fidelity susceptibility. 
Examples will be presented in Sec.~\ref{sec:BHM} and Sec.~\ref{sec:FHM}.


\subsubsection{Stochastic series expansion\label{sec:sse}}
\begin{figure}[t]
\includegraphics[width=9cm]{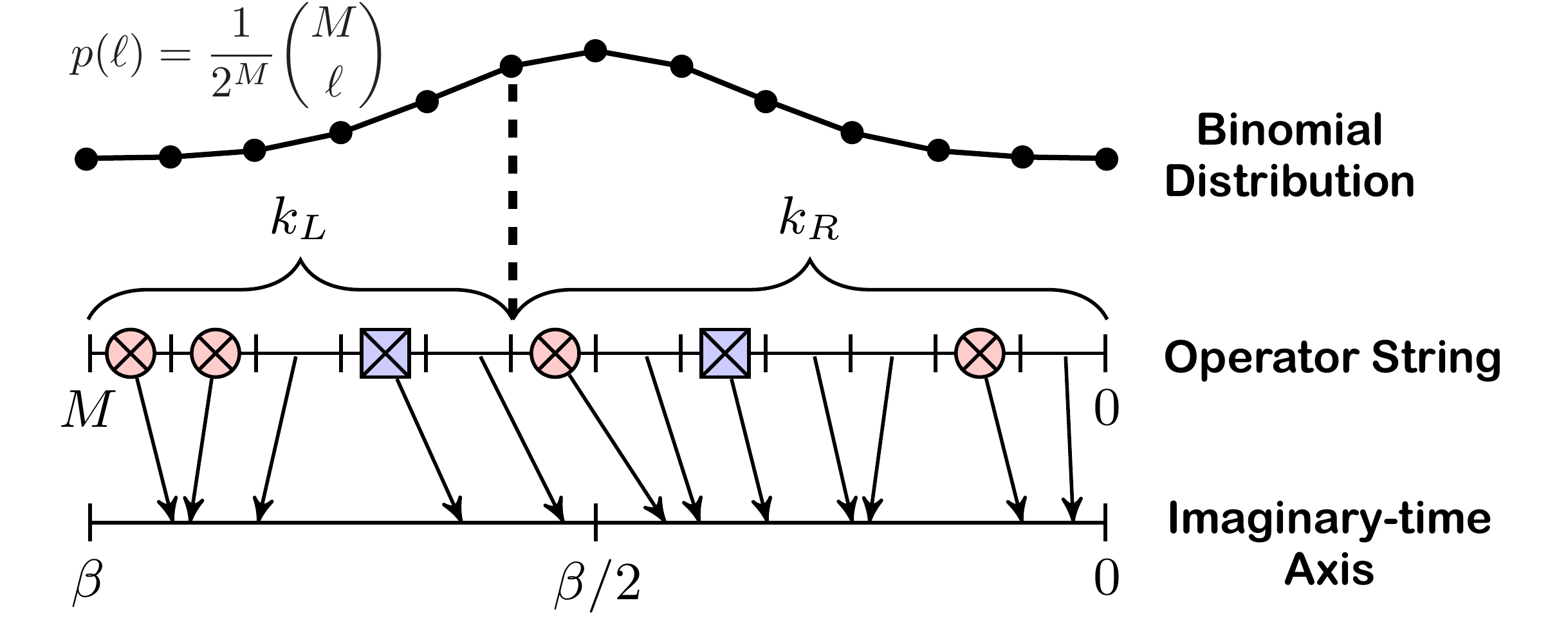}
\caption{Division of the operator string in SSE to measure the fidelity susceptibility. The slots represent the fixed-length operator string where empty slots hold identity operators, the red circles (blue squares) correspond to the operators in $\hat{H}_{1}$ ($\hat{H}_{0}$). These operators can be mapped to a continuous-time configuration indicated by the arrows. A division can then be made on the imaginary-time axis, for example at $\beta/2$. An equivalent approach without explicit mapping to continuous-time is to divide the operator string at the location indicated by the vertical dashed line, where the integer $\ell$ is drawn from a binomial distribution. For the estimators Eqs.~(\ref{eq:finiteTkLkR},\ref{eq:zeroTkLkR}) one counts the number of red circles (operators in $\lambda \hat{H}_{1}$) in both sides for $k_{L}$ and $k_{R}$. In this example $M=12, n=6, \ell = 7$, and  $k_{L}=k_{R}=2$. }
\label{fig:sseconfig}
\end{figure}

SSE is based on a Taylor expansion of the partition function~\cite{Sandvik:1991tn},
\begin{equation}
Z= \sum_{n=0}^{\infty}\frac{ (-\beta)^{n}}{n!} \, \Tr\left[\hat{H}^{n}\right],
\label{eq:sseexpansion}
\end{equation}
which may seem to be different from the framework of \Eq{eq:ftexpansion}. However, as is shown in Ref.~\cite{Sandvik:1997fb}, one can formally treat the SSE as the time-dependent expansion~\Eq{eq:ct-expansion} with respect to the full Hamiltonian $\hat{H}=\hat{H}_{0}+\lambda \hat{H}_{1}$.

In implementation of SSE, one truncates the sum to a large number $M$ and pads $M-n$ identity operators in the square bracket of \Eq{eq:sseexpansion}. SSE then samples operators in the fixed-length operator string. To map to a Monte Carlo configuration in the continuous-time formalism, one can assign an imaginary-time to each operator as shown in the bottom of Fig.~\ref{fig:sseconfig}. As long as the mapping keeps the relative order in the original operator string, the Monte Carlo weight remains unchanged~\footnote{In practice, one can generate imaginary-times randomly in the range of $[0, \beta)$, sort them in an ascending order and assign each one to an operator}. In particular, the configuration is sampled with a weight proportional to $\lambda^{k}$ if there are $k$ of $\lambda\hat{H}_{1}$ operators in the operator string. In this way, although the sampling of SSE is carried out differently from \Eq{eq:ct-expansion}, the general framework of Eq.~(\ref{eq:ftexpansion}) still applies. The fidelity susceptibility is then measured easily by counting the numbers $k_{L}$ and $k_{R}$ of operators associated with $\lambda\hat{H}_{1}$ in the two halves of the imaginary time axis after the mapping.

From Fig.~\ref{fig:sseconfig} it is clear that even though one performs an equal bipartition in the imaginary-time axis, the corresponding location of division is not always in the center of the operator string. In fact, it is easier to directly sample the location of division in the operator string, as shown in the upper part of Fig.~\ref{fig:sseconfig}. A division at the $\ell$-th position ($\ell =0,1,\ldots,M$) means that  there are $\ell$ slots being mapped to one half of the imaginary-time axis and $M-\ell$ slots to the other half. Therefore the division $\ell$ itself follows a binomial distribution $p(\ell) = \frac{1}{2^{M}}\binom{M}{\ell}$ which can be sampled directly. In this way, the fidelity susceptibility can be efficiently calculated in SSE similar to the continuous-time QMC approaches discussed in Sec.~\ref{sec:ctqmc}. As $M\rightarrow\infty$, the binomial distribution approaches to a delta function peaked at the center of the operator string. Only in this limit the position in the operator string can be directly interpreted as the imaginary-time and a bipartition of the operator string in the center will yield the correct result for the fidelity susceptibility.

%
%
%
%
%

\section{Derivations~\label{sec:proof}}
In this section we derive the estimators for the fidelity susceptibility at nonzero temperature~\Eq{eq:finiteTkLkR} and for ground-state projector formalism~\Eq{eq:zeroTkLkR}. Readers may skip this section and continue reading with the following sections. 

\subsection{Ground state}
%
%
%
Using the diagrammatic expansion for the projection operator, the unnormalized ground-state wavefunction \Eq{eq:gs} has the following form,

\begin{eqnarray}
\ket{\Psi_{0}} & = &\lim_{\beta\rightarrow\infty} \sum_{k=0}^{\infty}\lambda^{k} \int_0^{\beta/2}d  \tau_1 \ldots \int_{\tau_{k - 1}}^{\beta/2}d \tau_k \times \label{eq:projection} \\  &&\left[(-1)^{k}e^{- (\beta/2 - \tau_k) \hat{H}_{0}} \hat{H}_{1}  \ldots \hat{H}_{1}e^{- \tau_1 \hat{H}_{0}} \right] \ket{\Psi_{T}}. \nonumber
\end{eqnarray}
Substituting this into \Eq{eq:differentialform}, one obtains \Eq{eq:zeroTkLkR}. The estimator also holds for the continuous-time auxiliary field expansion methods (CT-AUX~\cite{Gull:2008cm} and LCT-AUX~\cite{Iazzi:2014vv}), because one can cast the ground-state wavefunction to a similar form as \Eq{eq:projection}, assuming the shift parameter used in these methods~\cite{1999PhRvL..82.4155R} to be proportional to $\lambda$.

\subsection{Non-zero temperature}

We present two derivations of the non-zero temperature estimator \Eq{eq:finiteTkLkR}.

\subsubsection{Derivation based on the definition of non-zero temperature fidelity \label{sec:proofdef}}

The non-zero temperature estimator Eq.~(\ref{eq:finiteTkLkR}) can be obtained directly from the definition of the non-zero temperature fidelity~\footnote{It is different form the Uhlmann fidelity~\cite{Uhlmann:1976ti, Jozsa:1994ik}.},
\begin{equation}
 F  = \sqrt{  \frac{\Tr \left(e^{- \beta \hat{H}(\lambda) / 2} e^{- \beta \hat{H}(\lambda+\epsilon) /
  2}\right)}  {\left(\Tr (e^{- \beta \hat{H}(\lambda) } ) \Tr (e^{- \beta \hat{H}(\lambda+\epsilon)}) \right)^{1 / 2}} }.
  \label{eq:FfiniteT}
\end{equation}
This is a non-zero temperature generalization of \Eq{eq:gsfidelity} and leads to Eq.~(\ref{eq:kubo}) by using the definition of the fidelity susceptibility~\Eq{eq:definition}~\cite{Sirker:2010fu}. We expand the traces of the density matrices around the partition function  $Z= \Tr (e^{- \beta \hat{H}(\lambda) } ) $ to $ O(\epsilon^{2})$,
\begin{eqnarray}
\Tr\left( e^{- \beta H(\lambda+\epsilon)}\right)   &  =  & Z + \epsilon
\partial_{\lambda} Z+   \frac{\epsilon^{2}}{2} \partial^{2}_{\lambda}Z , \nonumber \\
\Tr\left(e^{- \beta \hat{H}(\lambda) / 2} e^{- \beta H(\lambda+\epsilon) / 2}\right)  & = &  Z +
  \epsilon \vec{\partial}_{\lambda} Z + \frac{\epsilon^2}{2}
  \vec{\partial}_{\lambda}^2 Z ,\nonumber
\end{eqnarray}
where the notation $\vec{\partial}_{\lambda} $ indicates that the partial derivative acts only on operators in the imaginary time interval $0\le\tau<\beta/2$. Substituting the above two expansions into \Eq{eq:FfiniteT} and keeping terms up to $O(\epsilon^{2})$, one obtains

\begin{eqnarray}
\chi^{T\neq 0}_{F}
&= &\frac{( \vec{\partial}_{\lambda} Z)^{2}}{2Z^{2}} -\frac{ \vec{\partial}_{\lambda}^2 Z }{2 Z}
 + \frac{\partial^2_{\lambda}Z }{4 Z}  - \frac{(\partial_{\lambda}Z )^{2}}{4 Z^2} \nonumber \\
   & = & \frac{\braket{k_{R}}^{2} }{2\lambda^{2}}- \frac{\braket{k_R (k_R - 1)}}{2\lambda^{2}} + \frac{\braket{ k (k - 1) }}{4\lambda^{2}} - \frac{\braket{ k
  }^2}{4\lambda^{2}} \nonumber \\
  & = & \frac{  \braket{ k_L k_R } -\braket{k_{L}}\braket{k_{R}} }{2\lambda^{2}}. 
 \end{eqnarray}
We have used the partition function in the form of  \Eq{eq:ftexpansion} to obtain the second line, and $\braket{k_{L}} = \braket{k_{R}} = \braket{k}/2$ for the third line. This derivation is abstract and is independent to the details of a QMC scheme. Carrying out a similar procedure starting from \Eq{eq:gsfidelity}, one can also prove the ground-state estimator \Eq{eq:zeroTkLkR}.

\subsubsection{Derivation based on the imaginary-time correlator Eq.~(\ref{eq:kubo})}

This derivation starts from the definition of fidelity susceptibility based on the imaginary-time correlator Eq.~(\ref{eq:kubo}). We utilize its connection to the Monte Carlo weight appeared in the \Eq{eq:ct-expansion} to derive the non-zero temperature estimator \Eq{eq:finiteTkLkR}.
First of all, the second term in the square bracket of Eq.~(\ref{eq:kubo}) can be measured directly from the average expansion order~\cite{Sandvik:1991tn, 1999PhRvL..82.4155R, Rubtsov:2005iw, Gull:2011jd},
\begin{equation}
\braket{\hat{H}_{1}} = -\frac{\braket{k}}{\beta\lambda}.
\label{eq:k}
\end{equation}
Integrating over the imaginary-time and use $\braket{k_{L}} = \braket{k_{R}} = \braket{k}/2$, one has 

\begin{equation}
\int_{0}^{\beta/2}d\tau \left[ -\braket{\hat{H}_{1}} ^{2} \right] \tau =-\frac{\braket{k}^{2}}{8\lambda^{2}} = -\frac{\braket{k_{L}}\braket{k_{R}}}{2\lambda^{2}}
\label{eq:secondterm}. 
\end{equation}

We then consider the QMC estimator for the first term of \Eq{eq:kubo}
\begin{eqnarray}
G\left(\tau_{1}-\tau_{2}\right) & \triangleq & \left\langle \mathcal{T}\left[\hat{H}_{1}\left(\tau_{1}\right)\hat{H}_{1}\left(\tau_{2}\right)\right]\right\rangle \nonumber \\
 & = & \frac{1}{\lambda^{2}}\left\langle \sum_{i\neq j}\delta\left(\tau_{i}-\tau_{1}\right)\delta\left(\tau_{j}-\tau_{2}\right)\right\rangle ,\label{eq:estimator of G}
\end{eqnarray}
where $\mathcal{T}$ is the time ordering operator, $\tau_{i}$ and $\tau_{j}$ are the imaginary times of two vertices in the Monte Carlo configuration. Integrating both sides of \Eq{eq:estimator of G} and using the fact that $G\left(\tau_{1}-\tau_{2}\right)$
only depends on $\left|\tau_{1}-\tau_{2}\right|$, one has
\begin{align}
\int_{0}^{\Lambda}d\tau_{1}\int_{0}^{\Lambda}d\tau_{2}G\left(\tau_{1}-\tau_{2}\right)&=&2\int_{0}^{\Lambda}d\tau G
\left(\tau\right)\left(\Lambda-\tau\right)\nonumber \\
&=&\frac{1}{\lambda^{2}}\left\langle k\left(\Lambda\right)\left[k\left(\Lambda\right)-1\right]\right\rangle,\label{eq:identity of G}
\end{align}
where $k\left(\Lambda\right)$ is the number of vertices in the range of $0\le\tau<\Lambda$. For example, $k\left(\beta\right)=k$ and $k\left(\beta/2\right)=k_{R}$. When choosing
$\Lambda=\beta$ and using $G\left(\tau\right)=G\left(\beta-\tau\right)$, Eq.~(\ref{eq:identity of G}) becomes
\begin{equation}
\beta\int_{0}^{\beta}d\tau G\left(\tau\right) = \frac{1}{\lambda^{2}}\left\langle k\left(k-1\right)\right\rangle. \label{eq:kk-1}
\end{equation}
When setting $\Lambda=\beta/2$, Eq.~(\ref{eq:identity of G})
reads
\begin{eqnarray}
2\int_{0}^{\beta/2}d\tau G\left(\tau\right)\left(\frac{\beta}{2}-\tau\right)=\frac{1}{\lambda^{2}}\left\langle k_{R}\left(k_{R}-1\right)\right\rangle.
\end{eqnarray}
Together with \Eq{eq:kk-1} it leads to
\begin{eqnarray}
\int_{0}^{\beta/2}d\tau G\left(\tau\right)\tau & = & \frac{\left\langle k\left(k-1\right)\right\rangle }{4\lambda^{2}}-\frac{\left\langle k_{R}\left(k_{R}-1\right)\right\rangle }{2\lambda^{2}}\nonumber \\
 & = & \frac{\left\langle k_{L}k_{R}\right\rangle }{2\lambda^{2}}.
 \label{eq:GkLkR}
\end{eqnarray}
In combination with \Eq{eq:secondterm} we have derived Eq.~(\ref{eq:finiteTkLkR}).

\section{Applications~\label{sec:application}}
We first demonstrate the power of the new approach by identifying quantum and classical phase transitions in the Bose-Hubbard model and in the spin-$1/2$ XXZ model. Then we use the fidelity susceptibility to address the presence of the intermediate quantum spin liquid state in the Hubbard model on the honeycomb lattice. In all cases this required only minimal modifications to existing codes. We have purposely chosen a variety of QMC methods in the following to demonstrate the wide applicability of the covariance estimators. Because of the flexibility of the non-zero temperature estimator we used \Eq{eq:finiteTkLkR}. In Sec.~\ref{sec:factor2} we compare it to the zero-temperature scheme.

\subsection{Quantum phase transition in the Bose-Hubbard model \label{sec:BHM}}
First we use the fidelity susceptibility to probe the quantum phase transition in the Bose-Hubbard model,

\begin{figure}[b]
\includegraphics[width=9cm]{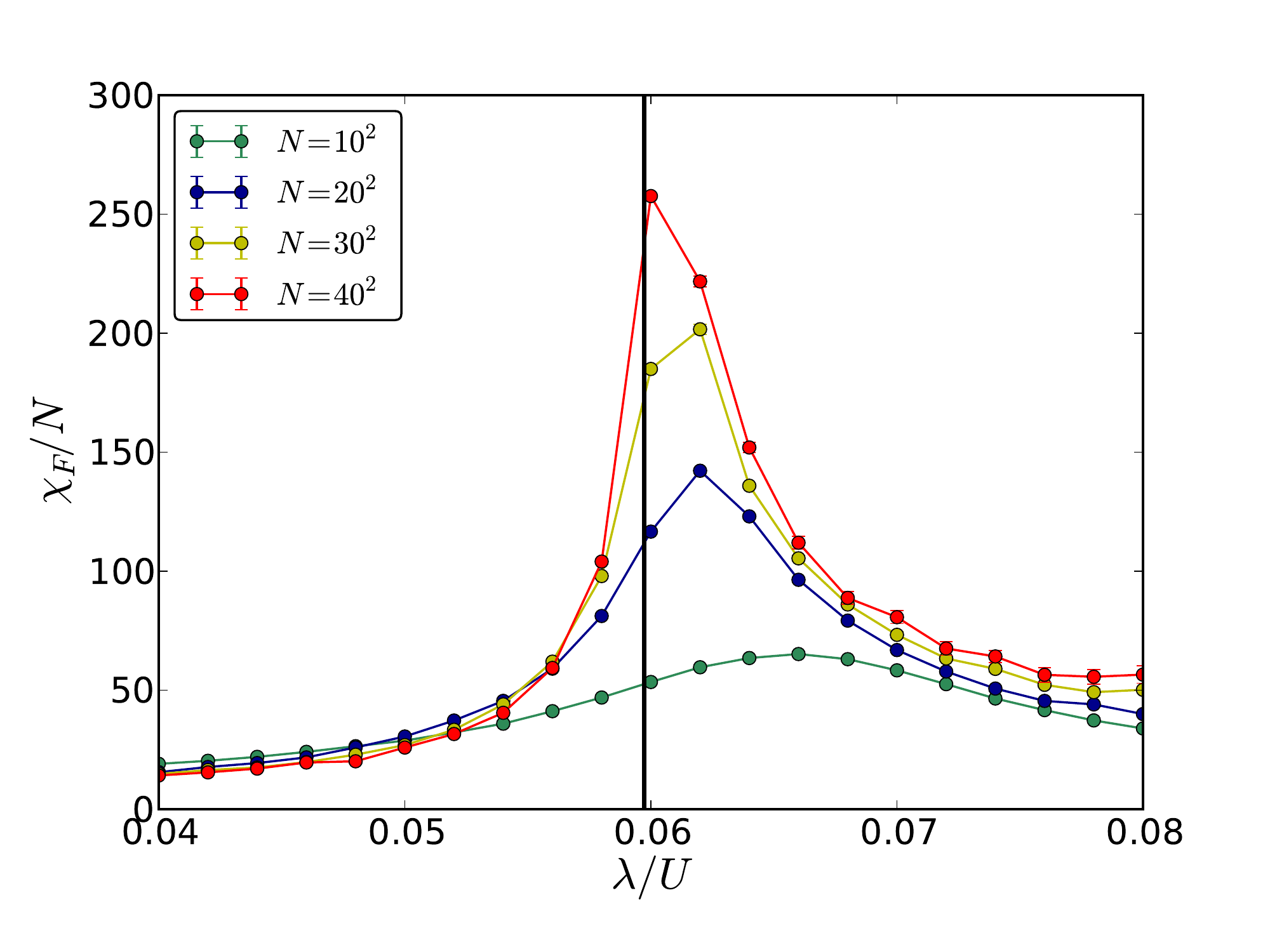}
\caption{Fidelity susceptibility per site of a Bose-Hubbard model on a square lattice at unit filling. The vertical line indicates the critical point determined in Ref.~\cite{CapogrossoSansone:2008eua}. }
\label{fig:BHM}
\end{figure}

\begin{align}
  \hat{H}  = \sum_{i}\left( \frac{U}{2} \hat{n}_{i} ( \hat{n}_{i} - 1) -\mu \hat{n}_{i}\right) -\lambda \sum_{\langle {i,j} \rangle}  \left( \hat{b}_{i}^{\dagger} \hat{b}_{j} + \hat{b}_{j}^{\dagger} \hat{b}_{i}\right), 
\end{align}
where $U$ is the on-site interaction and $\mu$ is the chemical potential. The driving parameter $\lambda$ has the physical meaning of a tunneling amplitude. The Bose-Hubbard model has a well-known quantum phase transition between the Mott insulating state and the superfluid state as $\lambda/U$ increases~\cite{Fisher:1989vw, Greiner:2002wt}. In particular, for integer fillings the system has an emergent Lorentz invariance at the critical point and the dynamical critical exponent is $z=1$~\cite{Sachdevbook}.

The fidelity susceptibility has previously been calculated using density-matrix-renormalization-group for the one-dimensional Bose-Hubbard model~\cite{Buonsante:2007dg, Carrasquilla:2013dga, Lacki:2014jia}. We now calculate the fidelity susceptibility on a square lattice with $N=L^{2}$ sites at unit filling by tuning $\mu$. In accordance with the dynamical critical exponent $z=1$, we scale the inverse temperature proportionally to the system length $\beta U =L$. The simulation employs the directed worm algorithm~\cite{Prokofev:1998tc, Pollet:2007gz, Tamathesis}. We utilize \Eq{eq:finiteTkLkR} to sample the fidelity susceptibility by counting the number of kinks in the worldline configuration, as illustrated in Fig.~\ref{fig:CTQMC}(a). Figure~\ref{fig:BHM} shows that as the system size increases, the peak in the fidelity susceptibility (as a function of the driving parameter $\lambda$) is becoming more pronounced and is shifting towards the previously determined critical point $(\lambda/U)_{c}=0.05974(3)$~\cite{CapogrossoSansone:2008eua}. The ability to calculate the fidelity susceptibility using the state-of-the-art directed worm algorithm~\cite{Prokofev:1998tc, Pollet:2007gz, Tamathesis} will greatly advance the study of quantum phase transitions of ultracold bosons. It is worth to point out the fidelity susceptibility is related to the quantity (kinetic-energy correlator) previously calculated  in the study of Higgs mode in a two-dimensional superfluid~\cite{Pollet:2012ila}.

\subsection{Classical phase transition in the XXZ model\label{sec:xxz}}

\begin{figure}[b]
\includegraphics[width=9cm]{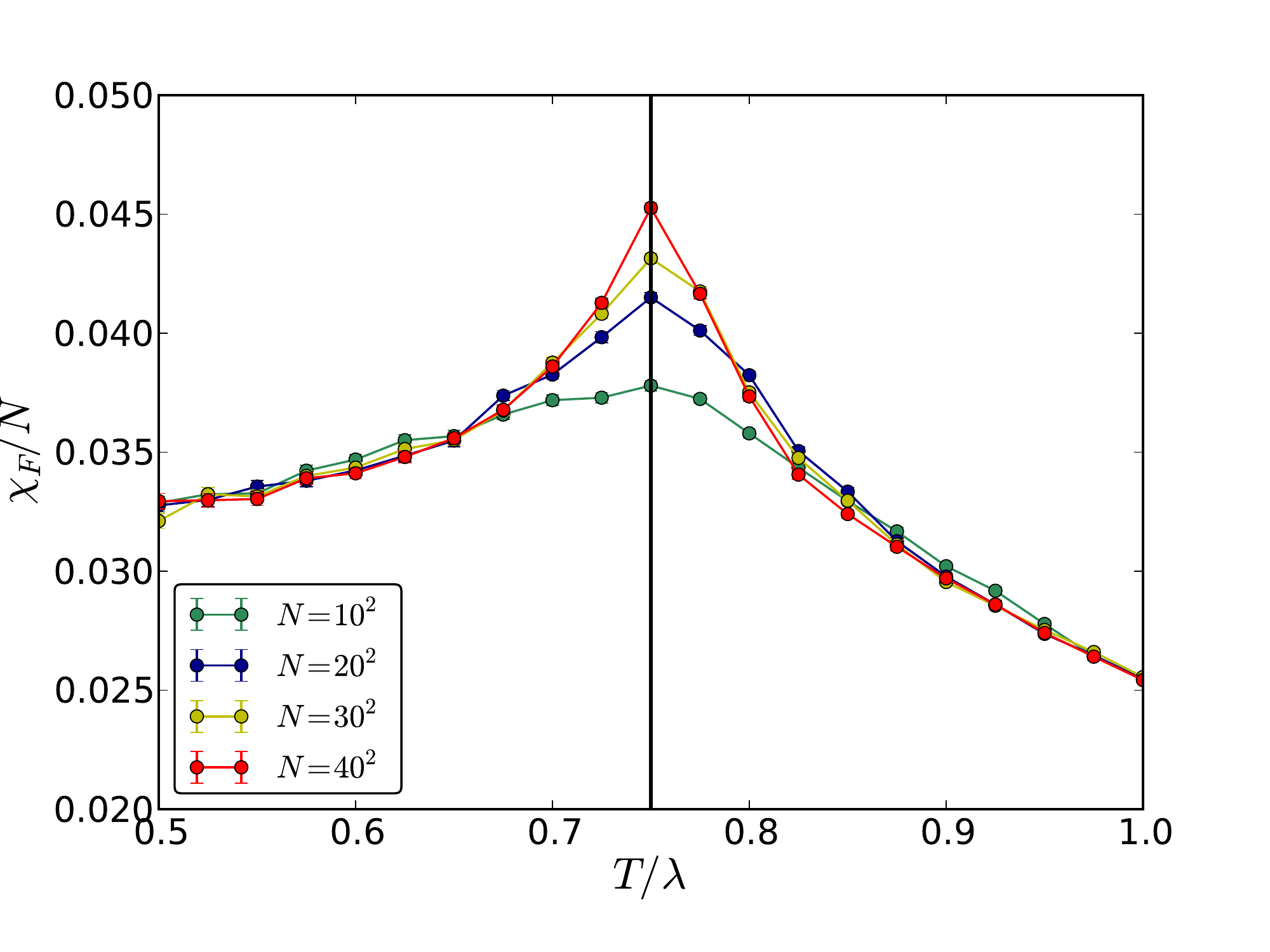}
\caption{Fidelity susceptibility per site of a XXZ model on square lattice versus temperature. The vertical line indicates the critical temperature determined in Ref.~\cite{Schmid:2002fu}. }
\label{fig:XXZ}
\end{figure}

Next we consider the spin-$1/2$ antiferromagnetic XXZ model on a square lattice with $N=L^{2}$ sites,
\begin{eqnarray}
  \hat{H} =  J_{z}\sum_{\langle{i,j}\rangle}\hat{S}^{z}_{i}\hat{S}^{z}_{j} +\lambda\sum_{\langle{i,j}\rangle}\left(\hat{S}^{x}_{i}\hat{S}^{x}_{j}+ \hat{S}^{y}_{i}\hat{S}^{y}_{j} \right),
\end{eqnarray}
where the driving parameter $\lambda$ plays the role of the coupling strength in the XY-plane. When $\lambda$ dominates the Hamiltonian favors N\'eel order in the XY-plane, while if $J_{z}$ dominates the system has an antiferromagnetic Ising ground state. The Heisenberg point $\lambda = J_{z}$ is a quantum critical point, which separates the XY order and the Ising order. This quantum critical point can be easily located from the peak of the fidelity susceptibility (not shown). Our approach makes it possible to obtain the fidelity susceptibility in much larger systems compared to the previous exact diagonalization study~\cite{PhysRevE.80.021108}, thus can enable a more accurate scaling analysis.

At nonzero temperature, thermal fluctuations will destroy the antiferromagnetic Ising phase at a second-order phase transition. Since one can cross the phase boundary either by changing $\lambda$ or the temperature $T$, we see that the fidelity susceptibility can also indicate thermal phase transitions. As a demonstration we fix $\lambda=1$, $J_{z}=1.5$ and scan the temperature $T$ to drive a phase transition from the low-temperature antiferromagnetic Ising phase to the high-temperature disordered phase. Figure~\ref{fig:XXZ} shows the fidelity susceptibility calculated using \Eq{eq:finiteTkLkR} via the SSE method~\cite{Sandvik:1991tn, Syljuasen:2002hw}. The peak in the fidelity susceptibility correctly single out the previously determined critical temperature $(T/\lambda)_{c} \approx 0.75$~\cite{Schmid:2002fu}.

\subsection{Intermediate phase in the Hubbard model on the honeycomb lattice\label{sec:FHM}}
Finally we apply the fidelity susceptibility estimator to a more challenging and controversial example -- the Hubbard model on the honeycomb lattice,
\begin{eqnarray}
  \hat{H} = &-&t\sum_{\langle {i,j} \rangle}\sum_{\sigma =\{\uparrow, \downarrow \}}  \left( \hat{c}_{i\sigma}^{\dagger} \hat{c}_{j\sigma} + \hat{c}_{j\sigma}^{\dagger} \hat{c}_{i\sigma}\right) \nonumber \\&+ & \lambda \sum_{i} \left( \hat{n}_{i\uparrow} - \frac{1}{2}
  \right) \left( \hat{n}_{i\downarrow} - \frac{1}{2} \right),
  \label{eq:Hubbard}
\end{eqnarray}
where $\lambda$ has the meaning of on-site Hubbard interaction strength. The simulation employs the recently developed efficient continuous-time QMC method for lattice fermions (LCT-INT)~\cite{Iazzi:2014vv}\footnote{In the practical simulation we use $\lambda<0$. However, because of the particle-hole symmetry, $\chi_{F}$ is symmetric around $\lambda=0$. Besides, we use $ \hat{H}_{1}  =  \sum_{i} [\left( \hat{n}_{i\uparrow} - \frac{1}{2}
  \right) \left( \hat{n}_{i\downarrow} - \frac{1}{2} \right) +\delta^{2}]$~\cite{Rubtsov:2005iw,Assaad:2007be}, where the constant shift $\delta=0.1$ ensures ergodicity of the Monte Carlo sampling. It has no effect on the results of $\chi_{F}$, but leads to a constant offset in Fig.~(\ref{fig:Gtau}).}. We consider lattices with $N=2L^{2}$ sites, with $L=6,9,12$ and scale the inverse temperature $\beta t=L$.

The ground-state phase diagram of the Hubbard model on the honeycomb lattice~\cite{Sorella:1992wd} has been controversial. It was suggested to possess an intermediate non-magnetic spin-liquid phase for $\lambda /t\in [3.5, 4.3]$~\cite{Meng:2010gc}. However, more recent QMC studies on larger systems~\cite{Sorella:2012hib} and with improved observables~\cite{Assaad:2013kg, Toldin:2014us} suggest a single continuous phase transition at $\lambda/t\approx3.8$ belonging to the Gross-Neveu universality class~\cite{Herbut:2006jaa}. Other less unbiased methods such as quantum cluster approaches give conflicting results on the presence of the intermediate phase~\cite{Yu:2011gf, Wu:2012bn, Hassan:2013kq, Liebsch:2013cw, Hassan:2013ei, Chen:2014jk}, depending on implementation details.

\begin{figure}[t]
\includegraphics[width=9cm]{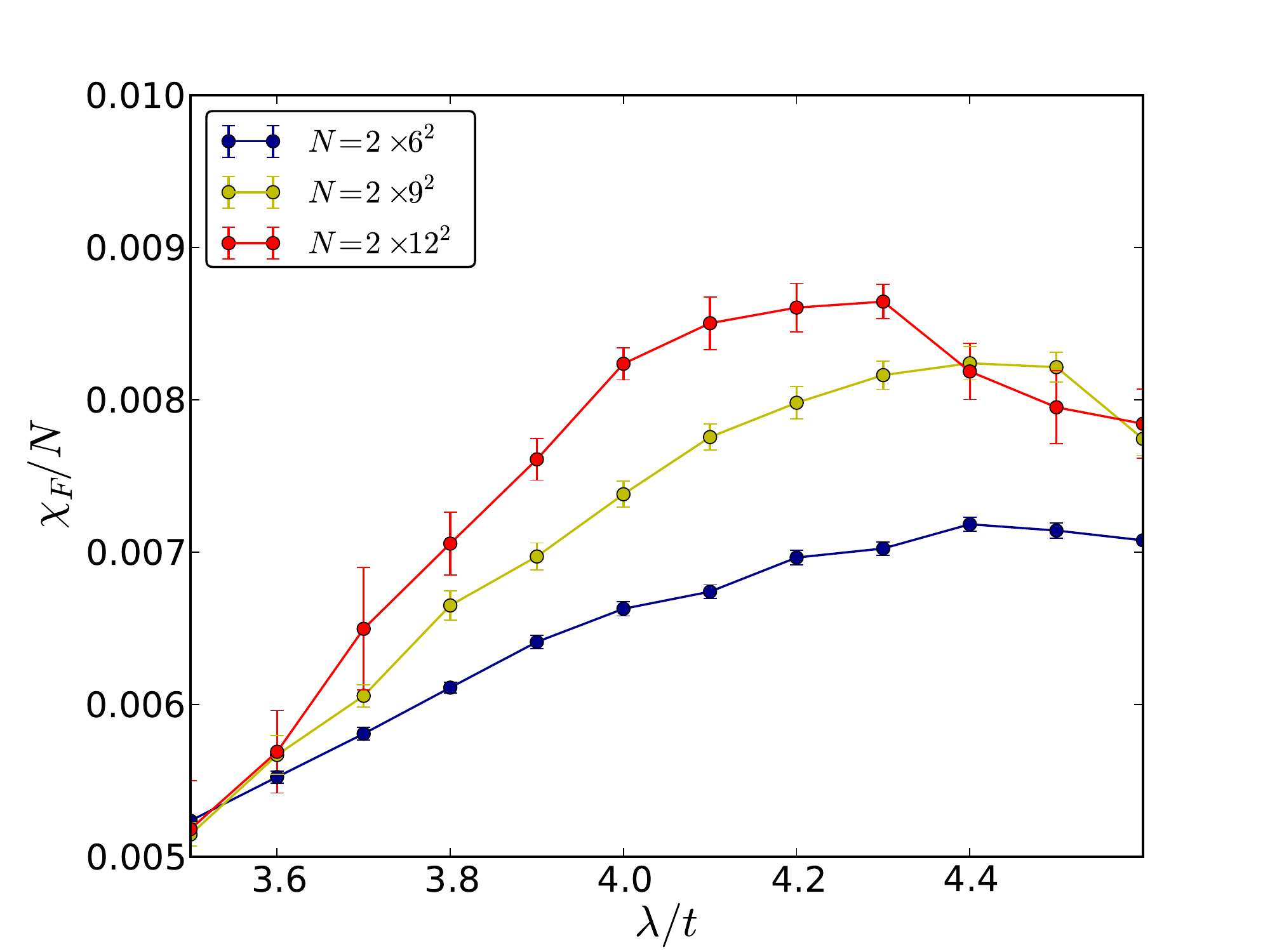}
\caption{Fidelity susceptibility per site 
of the Hubbard model on the honeycomb lattice Eq.~(\ref{eq:Hubbard}) with $N=2L^{2}$ sites.}
\label{fig:Hubbard}
\end{figure}

The fidelity susceptibility offers a new perspective on the debate about the phase diagram. In the scenario with an intermediate phase, there shall be \emph{two} features in $\chi_F$ when $\lambda /t$ approaches the two phase boundaries. This consideration is independent of the presence of a local-order-parameter description of the possible intermediate phase. Figure~\ref{fig:Hubbard} shows the fidelity susceptibility per site for various system sizes obtained using \Eq{eq:finiteTkLkR}. It exhibits a single broad peak for small systems (and high temperature). The peak becomes sharper and shifts towards smaller interaction strength as the system size increases. The fidelity susceptibility data presented in Fig.~\ref{fig:Hubbard} is consistent with a single phase transition at $\lambda/t\approx3.8$. In future studies for larger system sizes, and also with possible extension to a continuous range of $\lambda$ (by using histogram reweighting~\cite{1988PhRvL..61.2635F,Ferrenberg:1989tf} or quantum Wang-Landau approaches~\cite{Troyer:2003fta}), it may be possible to precisely determine the critical point, or even the critical exponent solely from the fidelity susceptibility data.


\section{Discussion~\label{sec:discussion}}

To help the reader gain a better understanding of the 
estimators \Eq{eq:finiteTkLkR} and \Eq{eq:zeroTkLkR}, 
we first discuss their relationship then compare them with the previous approach adopted in SSE calculations~\cite{Schwandt:2009jl, Albuquerque:2010fv}. Finally, we compare the fidelity susceptibility approach with other generic approaches for detecting phase transitions. 

\subsection{Relation of the ground-state and non-zero temperature estimators\label{sec:factor2}}

\begin{figure}[t]
\includegraphics[width=9cm]{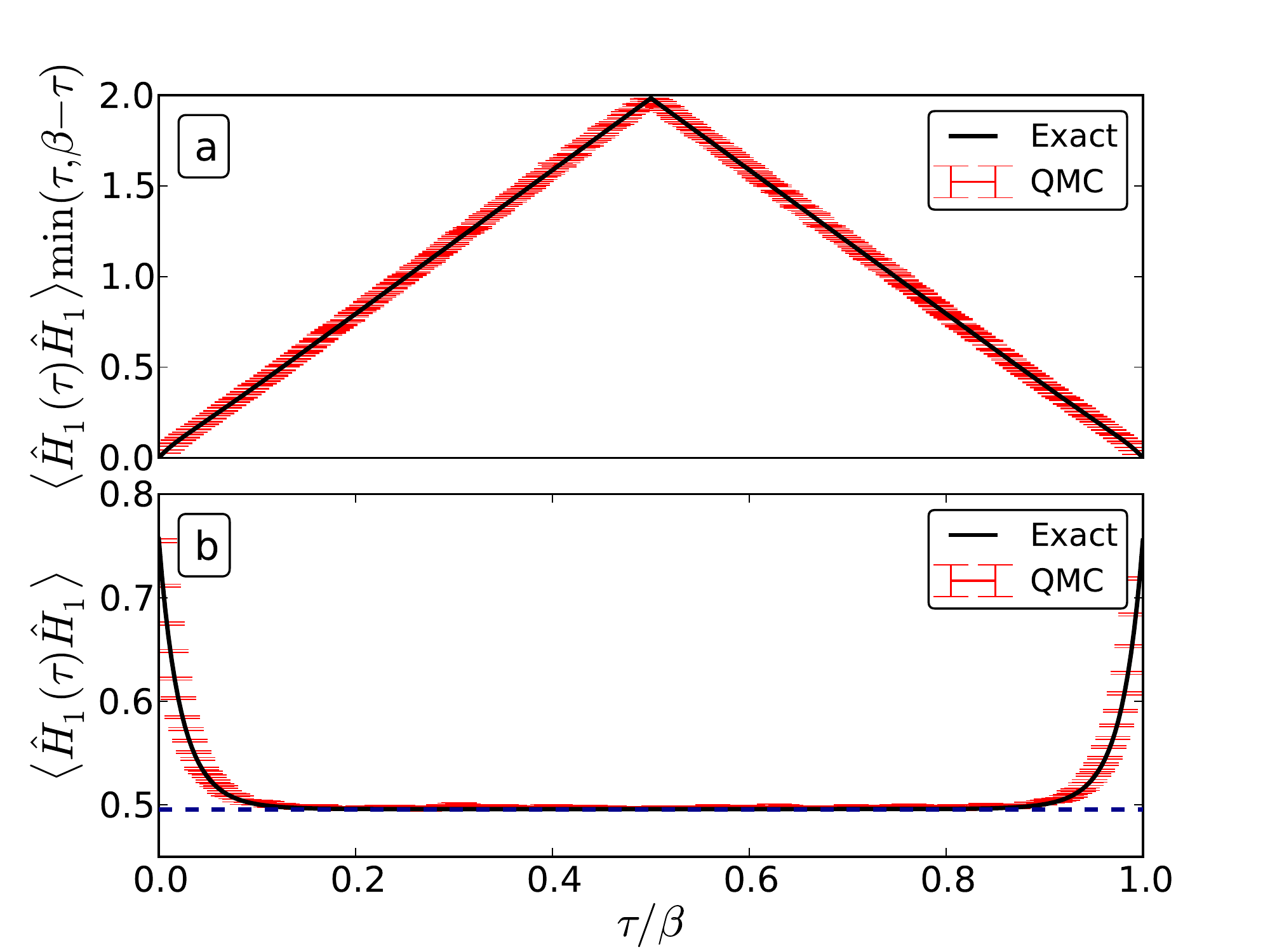}
\caption{(a) Histogram obtained by counting \emph{separable} vertex pairs with distance $\tau$ compared with the exact result of $\braket{\hat{H}_{1} \left({\tau} \right) \hat{H}_{1}}\times \max\{\tau, \beta-\tau\}$. (b) Histogram obtained by counting vertices with distance $\tau$ compared with the exact results of $\braket{\hat{H}_{1} \left({\tau} \right) \hat{H}_{1}}$. For $\tau$ close to $\beta/2$, the correlator approaches to $\braket{\hat{H}_{1}}^{2}$, indicated by the dashed blue line. These simulations are performed for the Hubbard model Eq.~(\ref{eq:Hubbard}) on a four-site open chain with $\lambda/t=-2$ and $\beta t=8$.}
\label{fig:Gtau}
\end{figure}

The factor of two difference in Eq.~(\ref{eq:finiteTkLkR}) and Eq.~(\ref{eq:zeroTkLkR}) is due to the different boundary conditions of the imaginary-time axis in the ground-state projection and non-zero temperature QMC formalisms, see Fig.~\ref{fig:concept}. We use a four-site Hubbard model (Eq.~(\ref{eq:Hubbard})) as an illustrative example. Consider the integrand of Eq.~(\ref{eq:kubo}), the correlator $G(\tau) = \braket{\hat{H}_{1} \left({\tau} \right) \hat{H}_{1}}$ is related to the distribution of the vertices on the imaginary-time axis. For a given configuration, the probability of finding two vertices with a time difference $\tau$ is proportional to $\lambda^{2}G(\tau)$. If we equally divide the imaginary-time axis into two halves and impose the additional constraint that the two vertices reside in different halves (denoted as a \emph{separable} vertex pair), the joint probability changes to $\lambda^{2}G(\tau) \min\{\tau, \beta-\tau\}$. Figure~\ref{fig:Gtau}(a) shows the histogram of \emph{separable} vertex pairs accumulated in the imaginary time, which indeed agrees with the exact curve. Summing up the histogram gives the total number of separable vertex pairs, which equals to the following integration,
\begin{equation}
\braket{k_{L} k_{R}}= \lambda^{2}\int_{0}^{\beta/2} d\tau G(\tau)\tau  +\lambda^{2}\int_{\beta/2}^{\beta} d\tau G(\tau)(\beta-\tau).
\label{eq:kLkR}
\end{equation}
Since $G(\tau)$ is symmetric around $\tau=\beta/2$ in the non-zero temperature simulation, the two terms of Eq.~(\ref{eq:kLkR}) are equal. Thus \Eq{eq:kLkR} reduces to \Eq{eq:GkLkR}. Furthermore, Fig.~\ref{fig:Gtau}(b) shows the correlator $G(\tau)$ sampled by accumulating the histograms of distances between vertices~\cite{Werner:2006ko, Gull:2011jd} together with the exact results (solid black line). The correlation between vertices decays rapidly with imaginary-time distance and approaches to the uncorrelated value $\braket{\hat{H}_{1}}^{2}$ (dashed blue line). 

\begin{figure}[t]
\includegraphics[width=8cm]{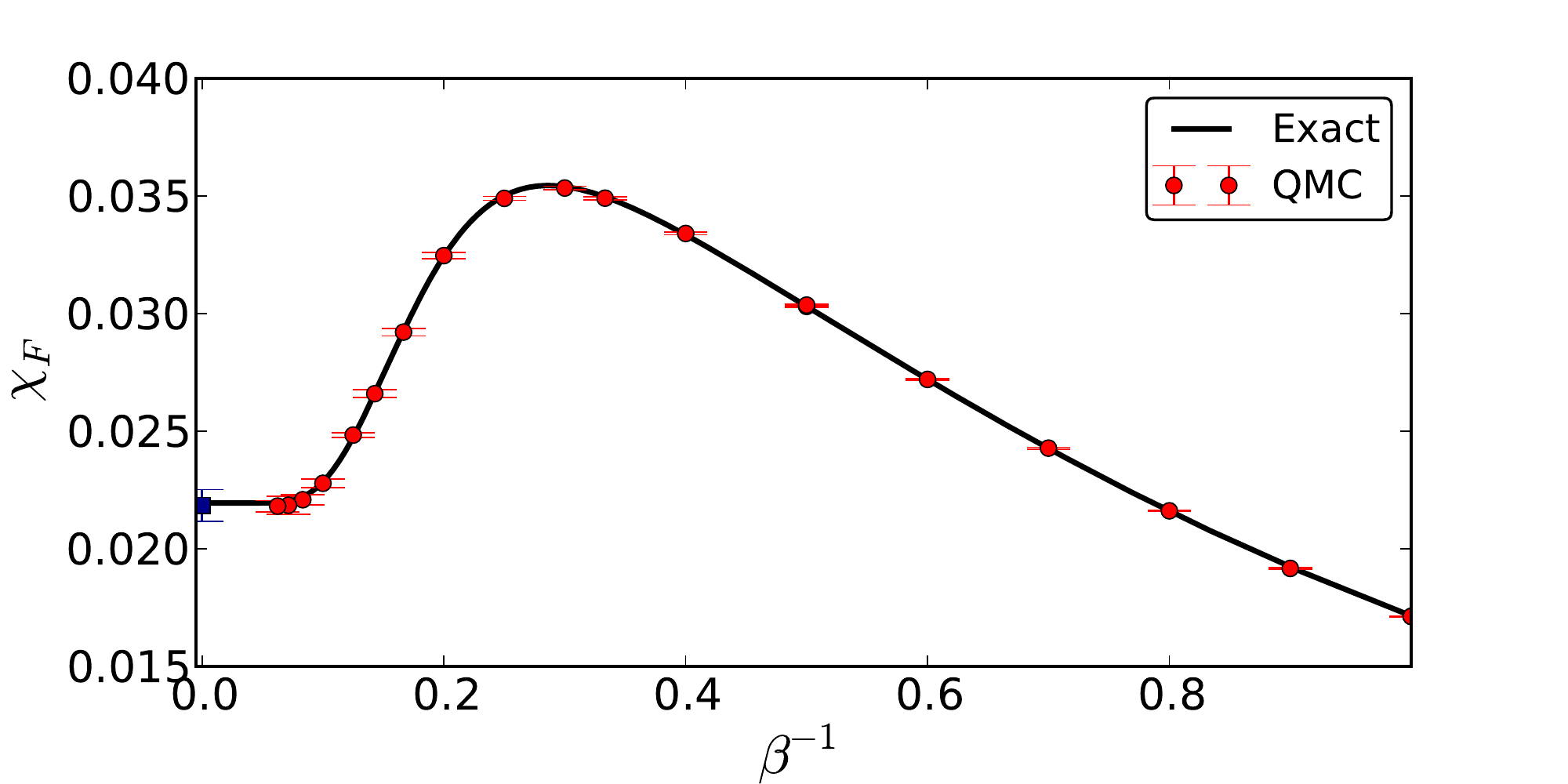}
\caption{QMC results for the fidelity susceptibility compared with exact results (solid line). The non-zero temperature QMC data (red dots) is obtained from Eq.~(\ref{eq:finiteTkLkR}), while the ground-state data (blue square at $\beta^{-1}=0$) is obtained from Eq.~(\ref{eq:zeroTkLkR}) in a projector LCT-INT calculation~\cite{Wang:2015tf}. The system is the same as in Fig.~(\ref{fig:Gtau}).}
\label{fig:benchmark}
\end{figure}

However, in the zero-temperature limit, the correlator $G(\tau)$ decays monotonically with $\tau$ and two vertices will decorrelate for $\tau\ge\beta/2$, where $\beta\rightarrow\infty$ in the projection scheme. Therefore the second term of Eq.~(\ref{eq:kLkR}) reduces to $\frac{\lambda^{2}\beta^{2}\braket{\hat{H}_{1}}^{2} }{8} = \frac{\braket{k}^{2}}{8}$ and cancels half of the second term in the estimator Eq.~(\ref{eq:zeroTkLkR}), resolving the apparent difference by the factor of two. In practical calculations, it is however crucial to adopt the correct formula to obtain consistent results, as illustrated in Fig.~\ref{fig:benchmark}. The fidelity susceptibility calculated using Eq.~(\ref{eq:finiteTkLkR}) in a non-zero temperature LCT-INT~\cite{Iazzi:2014vv} simulation agrees perfectly with exact diagonalization results. The blue square shows the value obtained using Eq.~(\ref{eq:zeroTkLkR}) in a projector LCT-INT calculation~\cite{Wang:2015tf}, which correctly reproduces the exact value of the ground-state fidelity susceptibility.

Figure~\ref{fig:Gtau}(b) also reveals the difficulty of computing the fidelity susceptibility. If the decay of $G(\tau)$ is faster than $1/\tau$, the integrand of Eq.~(\ref{eq:kubo}) has vanishing contributions at large $\tau$. However, as the two terms in Eq.~(\ref{eq:kubo}) are sampled independently in the actual QMC simulations, uncorrelated vertices at large imaginary-time distance will cause noises in the fidelity susceptibility signal. For the applications in Sec.~\ref{sec:application} we thus perform the calculations at nonzero temperature as it provides a natural cutoff. 

\subsection{Comparison to previous approaches}

The present approach to sample the fidelity susceptibility is more generic and efficient than those developed in Refs.~\cite{Schwandt:2009jl, Albuquerque:2010fv} specifically for the SSE method. It is nevertheless instructive to compare them in detail. The key difference lies in the sampling of the first term of \Eq{eq:kubo}. References~\cite{Schwandt:2009jl, Albuquerque:2010fv} employ the SSE estimator~\cite{Sandvik:1992wb}
\footnote{The original formula was derived for the reduced operator string, i.e. without padding the identity operators. However, the same formula holds as well for the fixed-length operator string.}
\begin{eqnarray}
G(\tau) &=& \frac{M-1}{\lambda^{2}\beta^{2}} \nonumber \\
  &\times& \sum_{n=0}^{M-2} \binom{M-2}{n}  \left(1-\frac{\tau}{\beta}\right)^{M-n-2} \left(\frac{\tau}{\beta}\right)^{n} \braket{  G(n) }, \nonumber
 \end{eqnarray}
where $G(n)$ is the number of occurrences of two operators from $\hat{H}_{1}$ that are separated by $n$ positions in the fixed-length operator string ($n=0$ if they are next to each other). Multiplying both sides with $ \max\{\tau, \beta-\tau\}$ and integrating over the imaginary time, one finds
\begin{align}
 \lambda^{2} \int_0^{\beta} d \tau\, G (\tau) \max\{\tau, \beta-\tau\}  =  \sum_{n = 0}^{M - 2} W(n) \braket{ G(n) } ,
 \label{eq:WG}
\end{align}
where the weight function $W(n)$ is written in terms of the regularized incomplete beta-function $I_{x}(a,b)$~\footnote{See, for example, the boost math library~\url{http://www.boost.org/doc/libs/1_57_0/libs/math/doc/html/math_toolkit/sf_beta/ibeta_function.html} for definition.},

\begin{eqnarray}
 W(n) &= & I_{\frac{1}{2}} (n+2, M-n -1) \frac{n+1}{M} +\nonumber \\& &  I_{\frac{1}{2}} (M - n, n + 1) \frac{M - n - 1}{M}.
 \label{eq:Wn}
 \end{eqnarray}

\begin{figure}
\includegraphics[width=8cm]{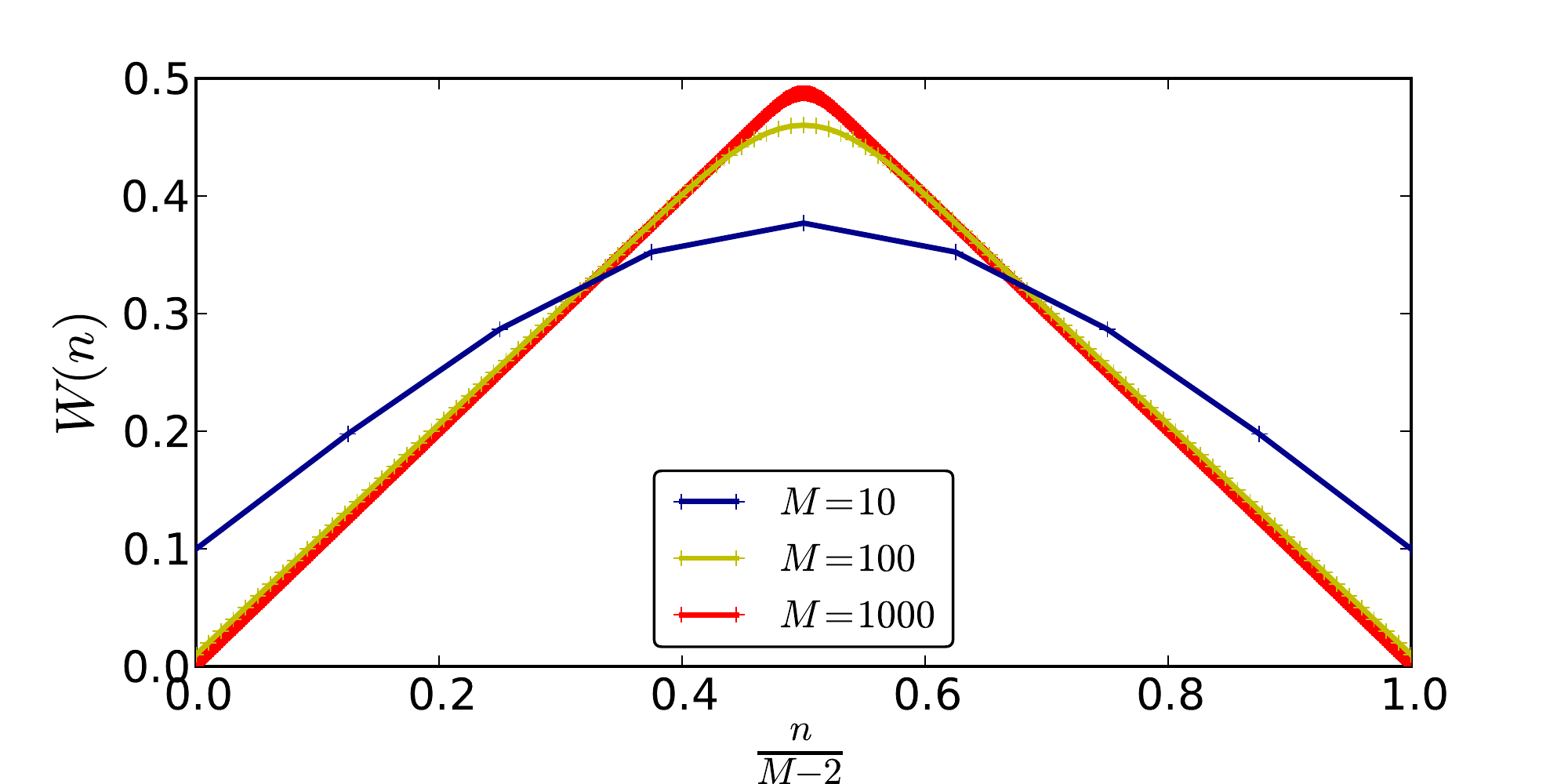}
\caption{The weight function according to \Eq{eq:Wn} for various truncation lengths $M$ in a SSE calculation.}
\label{fig:Wn}
\end{figure}
References~\cite{Schwandt:2009jl, Albuquerque:2010fv} \emph{explicitly} go through $k(k-1)/2$ pairs of vertices to accumulate $\braket{G(n)}$ and multiply it with the weight function $W(n)$. In \Eq{eq:kLkR}, however, the multiplication by the imaginary time $\tau$ is taken into account \emph{implicitly} by the sampling procedure (which requires separable vertices). Besides being more generic, our approach reduces the computational cost from $\mathcal{O}(k^{2})$ to $\mathcal{O}(k)$, which is crucial for the simulation of bosonic and quantum spin systems. In this sense, the specification of our general result \Eq{eq:finiteTkLkR} for the SSE method can be regarded as an improved estimator of \Eq{eq:WG}, which by itself already improves the approach of Refs.~\cite{Schwandt:2009jl, Albuquerque:2010fv} in several aspects~\footnote{Once the truncation $M$ is fixed in the SSE simulation (after equilibration), one can compute and store the \emph{one dimensional} array \Eq{eq:Wn}, where $I_{x}(a,b)$ is conveniently calculated using function calls to a numerical library.}. The improved estimator \Eq{eq:finiteTkLkR} not only unifies the SSE approach in a broader context of continuous-time diagrammatic QMC methods, it also gives better statistics with less computational cost compared to \Eq{eq:WG}.

Figure~\ref{fig:Wn} shows the weight function $W(n)$ for various truncation lengths. As $M$ increases, it approaches to two straight lines, and a division in the center of the operator string would yield increasingly accurate result for the fidelity susceptibility, consistent with the discussion in Sec.~\ref{sec:sse} concerning the large $M$ limit.



\subsection{Relationship to other quantities}
The fidelity susceptibility is related to the second-order derivative of the free energy $A  = -\frac{1}{\beta}\ln Z$ ~\cite{Chen:2008jr, Albuquerque:2010fv}. Due to the Hellmann-Feynman theorem~\cite{Hellmann1937, PhysRev.56.340}, $\braket{\hat{H}_{1}}$ equals to the first-order derivative of the free energy with respect to $\lambda$. A further derivative following the Kubo formula gives,

\begin{eqnarray}
\frac{\partial^{2} A}{\partial \lambda^{2}}=\frac{\partial \braket{\hat{H}_{1} }}{\partial \lambda} & = & -\int_0^{\beta} d\tau\, \left[\braket{\hat{H}_{1} \left({\tau} \right) \hat{H}_{1}} -  \braket{\hat{H}_{1}}^2\right] \nonumber \\
& = & \frac{\braket{k^{2}} - \braket{k} ^{2} - \braket{k}}{ -\beta \lambda^2}.
\label{eq:dDdU}
\end{eqnarray}
The third equality follows from \Eq{eq:k} and \Eq{eq:kk-1}~\footnote{The result can also be obtained directly by differentiating \Eq{eq:k} and using Eq.~(35) of Ref.~\cite{Wang:2015tf}.}. The quantity resembles the widely used SSE estimator for the specific heat~\cite{Sandvik:1991tn, Sengupta:2003ci}, but can be used to probe quantum phase transitions~\cite{Albuquerque:2010fv}. At zero temperature $\frac{\partial \braket{\hat{H}_{1} }}{\partial \lambda}= -\frac{1}{\lambda}\frac{\partial \braket{\hat{H}_{0}}}{\partial \lambda}$, and the latter quantity was computed using numerical differentiation of the kinetic energy, so as to address the quantum phase transition in the Hubbard model on the honeycomb lattice~\cite{Meng:2010gc, Chen:2014jk}. As is pointed out in Refs.~\cite{Chen:2008jr, Albuquerque:2010fv}, the fidelity susceptibility has a stronger singularity compared to the second-order derivative of the free energy and is thus a better indicator of quantum phase transitions. There are concrete examples in a class of topological phase transitions, which do not exhibit singularity in the second-order derivative of the ground-state energy~\cite{PhysRevB.78.035123}, but can still be detected using the fidelity susceptibility~\cite{Varney:2010eja}.

The covariance which appeared in the estimators \Eq{eq:finiteTkLkR} and \Eq{eq:zeroTkLkR} can also be written as
\begin{eqnarray}
\braket{k_{L} k_{R}} - \braket{k_{L}} \braket{k_{R}} = \frac{1}{2}\left[\mathrm{Var}(k)-\mathrm{Var}(k_{L}) - \mathrm{Var}(k_{R}) \right],  \nonumber
\end{eqnarray}
where $\mathrm{Var}(x) = \braket{x^{2}} - \braket{x}^{2} $ is the variance of $x$. This expression has an appealing meaning, i.e., the distributions of the vertices residing on the whole and on the halves of the imaginary-time axis have different widths, and the difference in these widths gives the estimate. This form resembles the bipartite fluctuation~\cite{Song:2012cp}, which was proposed to be a diagnostic tool for phase transitions~\cite{Rachel:2012eu} because of its relation to the entanglement entropy. However, there are important differences. First, the fidelity susceptibility estimator requires a division in the imaginary-time axis for vertices, not in the real space for the physical particles. Second, the total number of vertices is fluctuating in the QMC simulations as opposed to being conserved in the case of bipartition fluctuations. Third, it is easier to locate the critical point using the fidelity susceptibility. As is shown in this paper and in many previous studies~\cite{Gu:2010em}, the fidelity susceptibility exhibits an increasingly sharp peak at a phase transition as the system size enlarges. On the other hand, to utilize bipartite fluctuations and entanglement entropy for phase transition, one typically needs to resolve the scaling or subleading behavior with the system size, which is often difficult in finite size simulations.

\section{Outlook\label{sec:outlook}}
We have presented a general approach to compute the fidelity susceptibility of correlated fermions, bosons, and quantum spin systems in a broad class of quantum Monte Carlo methods~\cite{Sandvik:1991tn, PhysRevLett.77.5130, Prokofev:1998tc, 1999PhRvL..82.4155R, Evertz:2003ch, Kawashima:2004clb, Rubtsov:2005iw, Werner:2006ko, Burovski:2006hv, Gull:2008cm, Gull:2011jd, Iazzi:2014vv}. The calculation of the fidelity susceptibility is surprisingly simple yet generic. It provides a general purpose indicator of quantum phase transitions without the need for a prior knowledge of the local order parameter.

Conceptually, our work shows it is rewarding to view the modern QMC methods~\cite{Sandvik:1991tn, PhysRevLett.77.5130, Prokofev:1998tc, 1999PhRvL..82.4155R, Evertz:2003ch, Kawashima:2004clb, Rubtsov:2005iw, Werner:2006ko, Burovski:2006hv, Gull:2008cm, Gull:2011jd, Iazzi:2014vv} in a unified framework provided by \Eq{eq:ftexpansion}, which deals with the same type of classical statistical problem irrespective of microscopic details of the original quantum system. In the QMC simulations, a quantum phase transition manifests itself as a particle condensation transition driven by changing of the fugacity of the corresponding classical model. This connection suggests generic ways to detect and characterize quantum phase transition through studying classical particle condensations. For example, \Eq{eq:dDdU} actually relates the second-order derivative of free energy of a quantum system to the particle compressibility of a virtual classical system. In this respect, the significance of the covariance estimators Eqs.~(\ref{eq:finiteTkLkR},\ref{eq:zeroTkLkR}) is evident because they capture the key critical fluctuation upon a particle condensation transition. 


It is straightforward to generalize our Eqs.~(\ref{eq:finiteTkLkR},\ref{eq:zeroTkLkR}) to cases with multiple driving parameters, where one needs to count the vertices of different types (as is already done in the SSE calculations in Sec.~\ref{sec:sse} and Sec.~\ref{sec:xxz}). It is interesting to find out whether this can lead to a general approach to measure the Berry curvature (the imaginary part of the quantum geometric tensor) in quantum Monte Carlo simulations. Related to these efforts, the non-equilibrium QMC method is developed in recent years~\cite{Anonymous:2011kq, Liu:2013hb} to study non-adiabatic response of quantum systems in the imaginary time. In particular, it also allows the extraction of the fidelity susceptibility and the Berry curvature~\cite{Anonymous:2011kq, Kolodrubetz:2014kea}. It would be interesting to compare the non-equilibrium QMC approach~\cite{Anonymous:2011kq, Liu:2013hb} to the equilibrium one presented in this paper.

Last but not least, the Hamiltonian \Eq{eq:split} has further implications beyond quantum phase transitions. The efficient estimators Eqs.~(\ref{eq:finiteTkLkR}, \ref{eq:zeroTkLkR}) may also provide useful insights in the simulations of adiabatic quantum computation~\cite{Farhi:2001hy, Deng:2013dm, Boixo:2014cg} and non-adiabatic quantum dynamics~\cite{PhysRevB.81.012303, Polkovnikov:2011iu}.

\section{Acknowledgements}
The authors thank Mauro Iazzi, Sergei Isakov, Lode Pollet and Hiroshi Shinaoka for helpful discussions. Simulations were performed on the M\"{o}nch cluster of Platform for Advanced Scientific Computing (PASC) and on the Brutus cluster at ETH Zurich. We have used ALPS libraries~\cite{BBauer:2011tz} for Monte Carlo simulations and data analysis. The results of the Bose-Hubbard model were obtained using the \verb|dwa| (directed worm algorithm) application, and the results of the XXZ model were obtained using the \verb|dirloop_sse| (stochastic series expansion with directed loop updates) application of the ALPS project. This work was supported by ERC Advanced Grant SIMCOFE and by the Swiss National Science Foundation through the National Center of Competence in Research Quantum Science and Technology QSIT. 

\bibliographystyle{apsrev4-1}
\bibliography{FS_CTQMC}


\end{document}